\documentclass[a4paper,fleqn,usenatbib]{mnras}

\usepackage[T1]{fontenc}
\usepackage{ae,aecompl}

\usepackage{graphicx}	
\usepackage{amsmath}	
\usepackage{amssymb}	
\usepackage{tabularx,booktabs,tablefootnote}
\usepackage{multirow}
\usepackage{hyperref}
\usepackage{newtxtext,newtxmath}

\title[Jet of BP Psc]{Jet from the enigmatic high-latitude star BP Psc and evolutionary status of its driving source.}

\author[I. S. Potravnov et al.]{
Ilya S. Potravnov$^{1,2}$\thanks{E-mail: ilya.astro@gmail.com (ISP)},
M.Yu. Khovritchev $^{3}$,
S.A. Artemenko$^{4}$,
D.N. Shakhovskoy$^{4}$
\\
$^{1}$Institute of Solar-Terrestrial Physics, Siberian branch of the Russian Academy of Sciences, Lermontov Str. 126A, 664033, Irkutsk, Russia\\
$^{2}$Institute of Astronomy of the Russian Academy of Sciences, Pyatnitskaya str. 48, 119017, Moscow, Russia\\ 
$^{3}$Central (Pulkovo) Astronomical Observatory of the Russian Academy of Sciences, Pulkovskoye Chausse 65/1, 196140, St.Petersburg, Russia\\
$^{4}$Crimean Astrophysical Observatory of the Russian Academy of Sciences, p/o Nauchny, 298409, Republic of Crimea\\
}

\date{Accepted XXX. Received YYY; in original form ZZZ}

\pubyear{2022}

\begin{document}
\label{firstpage}
\pagerange{\pageref{firstpage}--\pageref{lastpage}}
\maketitle

\begin{abstract}
BP Psc is an active late-type (sp:G9) star with unclear evolutionary status lying at high galactic latitude $b=-57^{\circ}$. It is also the source of the well collimated bipolar jet. We present results of the proper motion and radial velocity study of BP Psc outflow based on the archival $H\alpha$ imaging with the GMOS camera at 8.1-m Gemini-North telescope as well as recent imaging and long-slit spectroscopy with the SCORPIO multi-mode focal reducer at 6-m BTA telescope of SAO RAS. The 3D kinematics of the jet revealed the full spatial velocity up to $\sim$140 km$\cdot$s$^{-1}$ and allows us to estimate the distance to BP Psc system as $D=135\pm40$ pc. This distance leads to an estimation of the central source luminosity $L_*\approx1.2L_{\odot}$, indicating that it is the $\approx$1.3$M_{\odot}$ T Tauri star with an age $t\lesssim$ 7 Myr. We measured the electron density of order $N_e\sim10^2$ cm$^{-3}$ and mean ionization fraction $f\approx0.04$ within the jet knots and estimated upper limit of the mass-loss rate in NE lobe as $\dot{M}_{out}\approx1.2\cdot10^{-8}M_{\odot}\cdot yr^{-1}$. The physical characteristics of the outflow are typical for the low-excitation YSO jets and consistent with the magnetocentrifugal mechanism of its launching and collimation. Prominent wiggling pattern revealed in $H\alpha$ images allowed us to suppose the existence of a secondary substellar companion in a non-coplanar orbit and estimate its most plausible mass as $M_p\approx 30M_{Jup}$. We conclude that BP Psc is one of the closest to the Sun young jet-driving systems and its origin is possibly related to the episode of star formation triggered by expanding supershells in Second Galactic quadrant.
\end{abstract}




\section{Introduction}
\defcitealias{Zuckerman2008}{Z08}

One of the most spectacular manifestations of the accretion/outflow activity of Pre-main-sequence (PMS) stars are the parsec-scale jets and Herbig-Haro flows \citep{Reipurth_2001}. Numerous jets and flows were discovered with the narrow-band imaging of the active star forming regions \citep[e.g.][]{Bally_2001}. According to the modern paradigm, they represent outer regions of the accretion-driven winds collimated by the large scale magnetic field \citep{Ferreira_2013,Ray_2021}. Jets of this type are the phenomenon of early stellar evolution and like their driving stars are genetically linked to natal giant molecular clouds and follow their spatial distribution. In the solar vicinity the scaleheight of the giant molecular clouds system is about $z_0\approx120$ pc or even less \citep[e.g.][]{Ferriere_2001,Stark_2005}. To an observer from the Earth this means that young stars and associated jets are predominantly inhabit the low-galactic latitudes: $|b|\lesssim25-30^{\circ}$. Therefore, the discovery of the collimated bipolar outflow driven by the star BP Psc \citep{Zuckerman2008}\citepalias{Zuckerman2008} located at high galactic latitude $b=-57^{\circ}$ about two decades ago was quite unexpected. It raised the questions about the evolutionary status of the system and mechanism that launched its jet.

Originally BP Psc was identified as an H$\alpha$-emitting object (StH$\alpha$ 202) by \citet{Stephenson1986} during the objective prism survey of high Galactic latitudes. The follow-up slit spectroscopy \citep{Downes&Keyes1988,Whitelok1995} revealed its late-type spectrum with superposed set of low excitation emission lines. The enhanced \ion{Li}{I} 6708\AA\ absorption which is characteristic feature of young T Tauri star (TTS) as well as \ion{He}{I} 5876\AA\, and \ion{O}{I} 8446 \AA\, emissions indicative for accretion activity were identified in high-resolution spectra of BP Psc by \citetalias{Zuckerman2008}. The latter authors estimated the mass accretion rate in the system as of order $\sim10^{-8}M_{\odot}\cdot yr^{-1}$. The extensive multiwavelength observations of BP Psc conducted by \citetalias{Zuckerman2008} also lead to a principal discoveries of the highly inclined ($i\approx70-80^{\circ}$) optically-thick circumstellar disk and collimated bipolar outflow.  

The bipolar outflow driven by BP Psc extends up to 6.5$\arcmin$ on either side from the source in the North-Eastern (NE) and South-Western (SW) directions at position angles PA$\approx24^\circ$ and PA$\approx210^\circ$ respectively. It shows the eastward curvature with the star in its apex and the brightness asymmetry between its NE and SW lobes. While SW lobe possesses weak [\ion{S}{II}] emission with single bright bow-shaped working surface emitting in H$\alpha$, the NE lobe is among the best examples of well-organized HH-flows with $H\alpha$-bright knotty filament close to the source and system of several terminal working surfaces at larger distances from the star. Such an arrangement of knots in the NE lobe looks very similar to the well investigated HH34 and HH30 young stellar jets \citep{Reipurth_1997,Anglada_2007}.

The PMS status of BP Psc was challenged by \citetalias{Zuckerman2008}, although the lithium overabundance, presence of an accretion disk of substantial mass and collimated jet seemed to support the classification of BP Psc as young T Tauri type star. Their main arguments against the star's youth were: (i) isolated position of BP Psc far from the active star-forming regions or young associations; (ii) weak ($EW\approx 50$m\AA\,) \ion{Li}{I} 6708\AA\ absorption which is not typical for still accreting TTS with age $t\lesssim$ 10 Myr ; (iii) low surface gravity $\log g\approx2.5$, estimated from the gravity-sensitive line ratios, which indicates luminosity class III rather than dwarf class IV-V expected for TTS. Alternatively \citetalias{Zuckerman2008} proposed the scenario with BP Psc being evolved star ascending the giant branch. Its disk and related phenomena in this scenario are the result of catastrophic accretion of the low-mass companion due to expansion of the primary.

However, in the absence of a known distance, and hence luminosity, the discussion of the evolutionary status of BP Psc had to rely on indirect proofs, which with some caveats could be interpreted within frameworks of both "young"\, and "old"\, scenarios. In any case this peculiar star deserves special attention. As was pointed by \citetalias{Zuckerman2008} if BP Psc belongs to TTS class, it is one of the closest to the Earth systems hosting an accretion disk, and is a prospective target for investigation with modern methods of high angular resolution. Otherwise, the mechanism of re-formation of the accretion disk and the launch of a jet in an evolved system is also of significant interest. With this in mind, our study was motivated by the possibility of obtaining an independent distance estimation and clarification of the evolutionary status of BP Psc, based on the investigation of the physical properties and 3D kinematics of its jet.

\section{Observations and data reduction}
To study jet kinematics for distance estimation, one needs data on the proper motion and radial velocity of the knots. For positional measurements of knots at two different epochs we combined both archival and our recent $H\alpha$ imaging data. We also obtained the long slit-spectroscopy across the outflow. The details of the observations are given below.  
\subsection{Imaging}
As the first epoch observations we used the images of BP Psc outflow retrieved from the Gemini Observatory Archive (obtained under the Program GN-2006B-DD-7, PI J. Graham). BP Psc was observed on the nights of 2007 January 18 and 23 with the 8.1-m Gemini-North telescope and \textsc{GMOS} camera. The nominal field of view was 2.6$\arcmin$ $\times$5.6$\arcmin$ and the plate scale was 0.145$\arcsec \cdot$px$^{-1}$. The images were obtained through the $H\alpha$ ($\lambda_0=6560$\AA\,, $\Delta \lambda$=70\AA\,) and continuum $g$ and $i$ filters. The resulted $H\alpha$ image was obtained by combining four frames with a total integration time of 2000$^s$. Observations were made under the excelent sub-arcsec seeing conditions.

Subsequently, BP Psc was observed on 2020 July 27-28 with the \textsc{SCORPIO} multi-mode focal reducer \citep{Afanasiev_2005} mounted in the prime focus of the 6-m BTA telescope of the SAO RAS (Russia). The observation strategy was to obtain a complete data set including imaging and spectroscopy for one of the NE and SW outflow lobes on a particular night. Observations were made with the EEV 42-40 CCD detector of 2048x2048 px size covering 6.1$\arcmin$ field of view. To increase the signal-to-noise ratio the 2x2 binning was applied which resulted in the actual plate scale of 0.356$\arcsec \cdot$px$^{-1}$. The frames were acquired through the $H\alpha$ filter with central wavelength $\lambda_0=6555$\AA\, and bandpass $\Delta \lambda$=75 \AA\,. The total exposure in the $H\alpha$ filter was 3000$^s$. We also obtained a 1200$^s$ exposure with the continuum SED707 ($\lambda_0=7036$\AA\,, $\Delta \lambda$=207\AA\,) filter. Due to unstable weather conditions and the relatively low altitude of the star above the horizon (observations were made at the air mass $X\approx$1.5), the resulting estimate of the seeing (from the point source) is $FWHM\lesssim2\arcsec$ for the 2020.07.27 observations of NE lobe and slightly worse for 2020.07.28 observations of SW one.  

The data processing of both data sets was made in the standard manner including the bias subtraction, flat-fielding and science frames combining.

The astrometric calibration of the 6-m BTA telescope images was performed using the \textsc{Gaia} EDR3 as a reference catalogue. All of the \textsc{Gaia} EDR3 reference star images were centered using the shapelet method \citep{Refregier_2003,Khovrichev_2018}. A model with the linear terms and cubic distortion terms was applied in the astrometric reduction procedure. The astrometric accuracy of about 50 mas was achieved. All of the $H\alpha$ images of the BP Psc field taken with the 8.1-m Gemini-North telescope were transformed into the  6-m BTA telescope image system using the results of the astrometric measurements of common sources (stars and galaxies). This allows further calculation of the knots' proper motion.

\subsection{Slit spectroscopy}

The optical spectra of BP Psc jet were obtained in the long-slit mode of SCORPIO reducer with the VPGH1200R grism which provides resolution $\delta \lambda=$5\AA\, in the 5750-7500\AA\, wavelength range. The same EEV 42-40 CCD was used with the 2x binning along only the spatial axis. This kept the spectral resolution undegraded and resulted in a similar 0.356$\arcsec \cdot$px$^{-1}$ image scale along the slit. The slit with 1.2$\arcsec \times$6$\arcmin$ dimensions in projection on the sky during the two observational runs was aligned first with the NE part of the outflow at position angle PA=24$^\circ$ and at PA=210$^\circ$ for the SW one. Total exposure time was 4500$^s$ and 1800$^s$ for NE and SW lobes respectively. Due to the curvature of the SW outflow, at this orientation of the spectrograph's slit it crossed only regions close to the source and terminal working surface, leaving the intermediate filament (S3, see below) unobserved. 

The spectroscopic data were processed with the \textsc{longslit} context of \textsc{IRAF} software. The scientific frames were corrected with both bias and normalized flat-field and after that combined to the master frame. Except for this standard calibration routine, special efforts were spent to correct the geometric distortion along the slit introduced by the SCORPIO optics. We followed a common approach where the required geometric transformation was determined from the spectrum of the He-Ne-Ar arc lamp and then applied to scientific data. The resulting quality of the transformation and wavelength calibration was checked by extracting the night-sky spectrum for several positions along the slit and comparing the measured positions of the emission features with the standard ones \citep{Osterbrock_1996}. The instrumental corrections for each spectral bin were determined from this comparison and then applied to the scientific data. Usually, their value did not exceed $\Delta \lambda\approx0.2$\AA\, ($\approx10$ km$\cdot$s$^{-1}$ in velocity scale). As was shown by \citet{Moiseev_2008} the largest errors are introduced in the measurements of radial velocities with SCORPIO by the effects of nonuniform slit illumination and differential refraction. The first effect should not affect much our observations since we observed an extended source and the angular size of the jet knots ($\beta \gtrsim 3\arcsec$) typically exceeds the slit width. Our data could be affected by differential refraction, since the orientation of the slit deviated from the paralactic angle due to the limitations of the observational strategy. However, the measured emission lines are distributed compactly within $\sim180$\AA\, spectral interval in the red part of the spectrum which should smooth out the effect of differential refraction on the relative measurements. Therefore, we estimate the resulting accuracy of our radial velocity measurements as $\pm$15 km$\cdot$s$^{-1}$, including in the error budget also the quality of the emission profile fitting.

\begin{table*}
\caption{Proper motions and radial velocities of knots in BP Psc jet}
\label{tab1}
\begin{tabular}{lcccccccccc}
\hline
Knot & RA & DEC & Separation & $\mu_{\alpha}$ & $\mu_{\delta}$  & $\mu$ & PA & $RV(H\alpha)$ & $RV[\ion{N}{II}]_{6548, 6583}$ &$RV[\ion{S}{II}]_{6716, 6731}$  \\
& (deg) & (deg) & ($\arcsec$) & ($\arcsec \cdot$yr$^{-1}$) & ($\arcsec \cdot$yr$^{-1}$) & ($\arcsec \cdot$yr$^{-1}$) & (deg) & (km$\cdot$s$^{-1}$) & (km$\cdot$s$^{-1}$) & (km$\cdot$s$^{-1}$) \\
\hline
NE lobe&&&&&&&&&&\\
N1-A & 350.6165 & -2.1988 &117 & 0.061  & 0.164 & 0.175 & 20.4 &+38 & +41 & +40  \\
N1-B & 350.6175 & -2.1968 &125 & 0.089 & 0.171 & 0.193 & 27.6 &+43 & +47 & +43 \\
N1-C & 350.6185 & -2.1949 &132 & 0.050 & 0.122 & 0.132 & 22.4 &+40 & +43 & +37  \\
N1-D & 350.6194 & -2.1929 &140 & 0.062 & 0.156 & 0.168 & 28.8 &+20 & +31 & +21  \\
N1-E & 350.6207 & -2.1903 &151 & 0.084 & 0.175 & 0.194 & 25.5 &+19 & +29 & +28  \\
N1-F & 350.6211 & -2.1881 &158 & 0.102 & 0.174 & 0.202 & 30.5 &+26 & +40 & +38 \\
N1-G & 350.6210 & -2.1876 &160 & 0.074 & 0.151 & 0.168 & 26.0 &... & ... & ...  \\
N2 & 350.6237 & -2.1833 &178 & 0.082 & 0.155 & 0.175 & 27.7 &+29 & +37 & +33  \\
N3 & 350.6251 & -2.1786 &196 & 0.062 & 0.115 & 0.131 & 28.6 &+20 & +41 & +38 \\
N4 & 350.6314 & 2.1686 &238 & 0.069 & 0.122 & 0.140 & 29.4 &+4 & +12 & +17  \\
N5 & 350.6312 & -2.1635 &254 & 0.037 & 0.066 & 0.076 & 29.7 &+6 & ... & ...  \\
\hline
SE lobe&&&&&&&&&&\\
S4 & 350.5797 & -2.2676 & 165 & 0.056 & -0.015 & 0.058 & 255 &-23 & ... & -31 \\
\hline
\end{tabular}
\end{table*}

\section{Results}

\subsection{Jet morphology}

\begin{figure*}
	\includegraphics[width=1.0\linewidth]{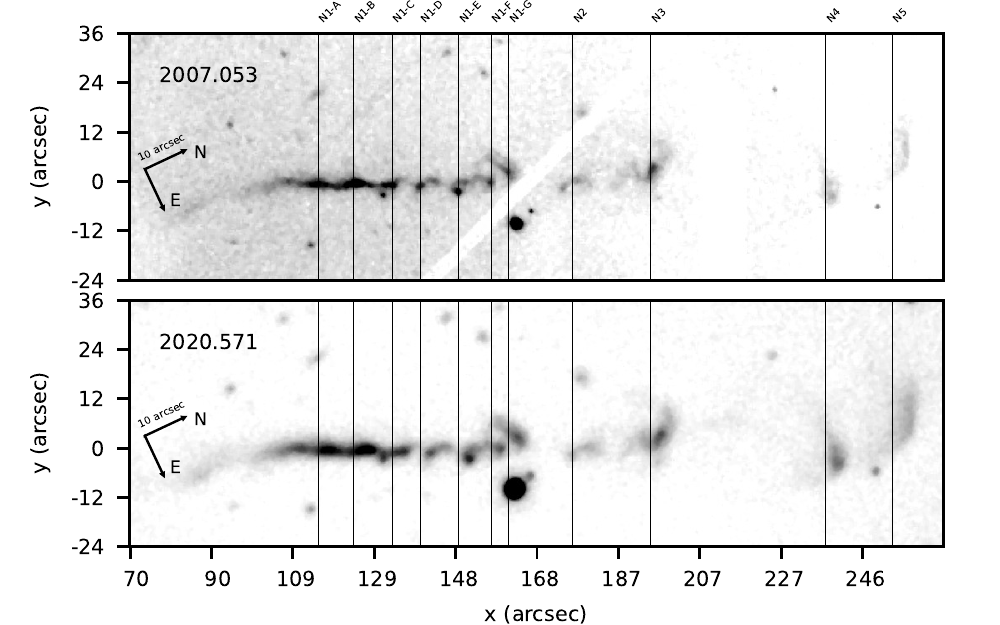}
    \caption{\textsc{GMOS} (top) and \textsc{SCORPIO} (bottom) $H\alpha$ images of NE lobe of BP Psc jet obtained with 13.5 years epoch difference. The origin of the coordinates coincides with the central source, which is not shown in the figure. The positions of the knots are marked with thin lines and signed at the top of the figure. }
    \label{fig:1}
\end{figure*}

The \textsc{GMOS} and \textsc{SCORPIO} narrow-band $H\alpha$ images we analyzed cover up to 5$\arcmin$ on either side from the star, thus the most distant faint knots N6-7 and S5-6 were outside of our field of view. Hereafter we follow the nomenclature of the knots from \citetalias{Zuckerman2008}. However, within the northern filament N1 we additionaly marked several well-defined sub-condensations N1-A-G which are suitable for positional and radial velocity measurements. A list of the examined knots with their coordinates and distances from the source is given in Tab.\ref{tab1}.

The NE outflow begins at the distance of about 1.4$\arcmin$ from the star as the bright chain N1 of HH objects. While N1-A-C is a continuous filament with a somewhat clumpy structure, the N1-D-G are well-detached knots with bright condensed nuclei. These star-like nuclei are most clearly visible in the \textsc{GMOS} image (Fig.\ref{fig:1}) obtained under the excellent seeing conditions. Further away from the star the northern outflow continues with more diffuse knots N2-3. Within our field of view it terminates by the two bow-shaped working surfaces N4-5 at distances 3.9$\arcmin$ and 4.2$\arcmin$ respectively. In our $H\alpha$ images the opposite SW lobe is less pronounced and represented only by the vague traces of the S3 filament and bright compact working surface S4 separated by 2.8$\arcmin$ from the star.

The picture is different in [\ion{S}{II}] emission \citepalias[Fig.11 in][]{Zuckerman2008} where the S3 filament turns out to be well-defined and locates more or less symmetrically to the position of the N1 chain with respect to the star. The S1-2 filaments of about 20$\arcsec$ length from the continuum source also emit in [\ion{S}{II}] with no accompanying $H\alpha$ emission as seen in \citetalias[][]{Zuckerman2008} images and along the spatial axis of our longslit spectrogram. 

Limited by the availability of only $H\alpha$ images, we will further concentrate on the discussion of the fine structure of the NE lobe. It is strongly collimated with full opening angle $\phi \approx2.2\pm1^{\circ}$ at the first 2.5$\arcmin$ from the source according to our measurements with crosshair in the contour plot of the \textsc{GMOS} $H\alpha$ image. The arrangement of the knots leaves the impression of being twisted around the axis of the flow. Such a pattern is very similar to the so-called "wiggling jets"\, frequently observed in young stellar objects (YSOs) \citep[e.g.][]{Terquem_1999,Anglada_2007}. Over the entire observed length of the NE jet, we can estimate the full opening angle of this wiggling pattern as $\approx 5\pm1.5^{\circ}$ using the same measurement procedure.

\subsection{Physical conditions along the jet and mass-loss rate}

\begin{figure}
	\includegraphics[width=1.0\linewidth]{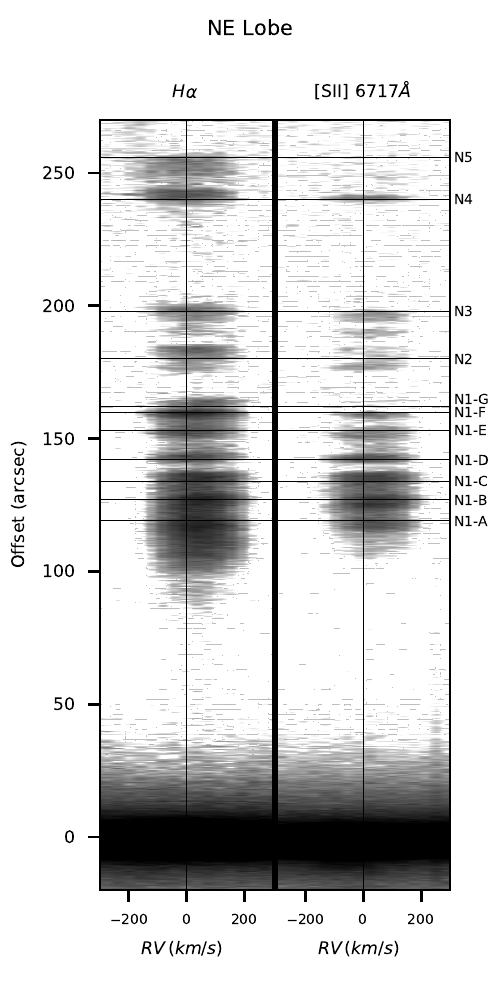}
    \caption{Position-velocity diagrams for NE lobe, plotted for $H\alpha$ (left panel) and [\ion{S}{II}] (right panel) lines. The positions corresponding to the knots in the $H\alpha$ image are marked with thin lines and signed on the right}
    \label{fig:2}
\end{figure}

Our spectrogram of BP Psc outflow revealed the principal emission lines observed in HH-jets, including $H\alpha$, [\ion{O}{I}] 6300, 6364\AA\,, [\ion{S}{II}] 6716, 6731\AA\,, and [\ion{N}{II}] 6548, 6583\AA\ doublets. Browsing along the spatial axis of the 2D spectra shows that the line ratios from knot to knot are variable and emission peaks of different transitions are spatially misaligned. Both of these effects are illustrated by the "position-velocity"\, (PV) diagram (Fig.\ref{fig:2}) constructed for $H\alpha$ and [\ion{S}{II}] lines. One can see that the NE outflow is brighter in $H\alpha$. In some individual knots, [\ion{S}{II}] emission is much weaker (e.g. N2) or even absent as in N1-G and N5. Where $H\alpha$ and [\ion{S}{II}] emissions appear simultaneously, the latter turns out to be spatially more compact and trailing the $H\alpha$-emitting region. 

In order to investigate the plasma conditions along the jet, we used the diagnostic line ratios. We extracted 1D spectra at several positions along the slit coincidental with the jet knots listed in Tab.\ref{tab1}. The spectra were binned along the spatial axis within the boundaries of these knots to integrate the signal. Thus, the measured line ratios refer to integral characteristic of the entire knot. Unfortunately, we were limited to ratios involving [\ion{S}{II}] and $H\alpha$ lines only, since measurements of the [\ion{O}{I}] 6300, 6364\AA\, lines proved to be unreliable due to the erratic extraction of strong night-sky emission component. Thus, we used the [\ion{S}{II}] $I_{6716}$/$I_{6731}$ as indicator of electron density $N_e$ and $(I_{6716}$+$I_{6731})$/$I_{H\alpha}$ ratio to measure shock velocities and ionization fraction using the diagrams computed for planar shock model by \citet{Hartigan_1994}. 

The measured line ratios and values of $N_e$ which were determined with the diagrams by \citet{Proxauf_2014} for $T_e=10^4K$ are summarized in Tab.\ref{tab2}. The resulting electron densities of order $N_e\sim10^2$ cm$^{-3}$ are typical for low-excitation YSO jets \citep[see e.g.][]{Hartigan_1994,Anglada_2007}. 

One can see that the largest $N_e$ values are observed closer to the source in the brightest parts of N1 filament. The lowest electron density was detected in knots N1-F, N2 and in bow-shocks N4, S4. These knots possess the lowest $I_{6716+6731}$/$I_{H\alpha}$ ratio, indicating higher shock velocity and, hence, high degree of ionization behind the shock. The shock velocities within the continuous part of NE outflow are $V_s\approx35-45$ km$\cdot$s$^{-1}$, and mean ionization fraction $f\approx0.04$. These values reach $V_s\approx50$ km$\cdot$s$^{-1}$ and $f\approx0.12$ for the stronger S4 bow-shock. However, as was pointed in \citet{Bacciotti_1999} due to the step temperature gradient behind the shock two mechanisms producing the $H\alpha$ emission operating simultaneously: collisional excitation in the compressed shock zone and recombination in the regions with lower temperatures. Inability to discriminate between two this mechanisms and improper accounting for the low-temperature recombination could lead to an underestimation of the ionization fraction by about 40$\%$ (for a intermediate-velocity shock, see their Fig.2), and hence an overestimation of the total hydrogen concentration. Therefore, one should keep in mind that the estimation of the mass-loss rate below resulted in the upper limit for the real value of $\dot{M}_{out}$.

To estimate the mass-loss rate we used the part of N1 filament confined by A and E knots. Following the approach by \citet{Hartigan_1994} we estimate the mean density $N\approx890$ cm$^{-3}$ assuming ionization fraction $f\approx0.04$ and compression factor $C\approx16$ according to the corresponding diagram for the mean value of shock velocity in this part of the outflow. The diameter of the outflow "tube"\, can be determined from its observed $\approx3.1\arcsec$ width, which at distance $D\approx135$ pc (see Sect.\ref{sec:Dist}) corresponds to 420 au. The average spatial velocity within this section of N1 filament $<V_{tot}>=\sqrt{V_r^2+(4.74\mu D)^2}\approx118$ km$\cdot$s$^{-1}$ was calculated from the proper motions and radial velocities (see Sect.\ref{sec:PM},\ref{sec:RV}) assuming inclination $i=75^{\circ}$. These quantities yield in the estimation of mass-loss rate in NE lobe as $\dot{M}_{out}\approx1.2\cdot10^{-8}M_{\odot}\cdot yr^{-1}$.

It should be emphasized that the abovementioned estimation of $\dot{M}_{out}$ refers only to the NE part of outflow. It is important to establish whether the mass-loss rate is the same in the SW lobe or not, in order to understand the reason for the difference in appearance of these two lobes. One is the real difference in $\dot{M}_{out}$, the other is that the SW lobe is less excited for some reasons. More extensive spectroscopy within the SE lobe, particularly covering the S3 filament, is needed to discriminate between these possibilities. 

 \begin{table}
\caption{Observed line ratios and electron densities in BP Psc jet}
\label{tab2}
\begin{tabular}{lccc}
\hline
Knot & $I_{6716}$/$I_{6731}$ &  $I_{6716+6731}$/$I_{H\alpha}$ & $N_e$, cm$^{-3}$\\  
\hline
NE lobe&&&\\
N1-A & 1.27 & 0.88  & 138\\
N1-B & 1.20 & 1.30  & 204\\
N1-C & 1.10 & 1.34  & 319\\
N1-D & 1.36 & 1.53  & 74\\
N1-E & 1.29 & 1.25  &122\\
N1-F & 1.40 & 0.60  & 53\\
N2 & 1.51 & 0.55    & 18\\
N3 & 1.19 & 0.94    & 214\\
N4 & 1.31 & 0.62    & 106\\
\hline
SW lobe&&&\\
S2 & 1.19 & ...  & 214  \\
S4 & 1.40 & 0.47 & 53\\
\hline
\end{tabular}
\end{table}

 Curiously, we detected an unidentified emission line in both lobes of BP Psc outflow. The moderately bright emission at $\lambda$6473.86\AA\ was observed at 4$\arcmin$ distance from the star in the SW beam and has no counterparts in other emission lines. In the NE beam it can be tentatively attributed to the trailing side of the N4 working surface. The closest match in wavelength corresponds to the [\ion{Fe}{II}] 6473.86\AA\, line of the 44F multiplet (according to Moore's tables) arising from $^2G-^2F$ transition. As far as we know, this line is not typical for low-excitation jets but is sometimes observed in spectra of planetary nebulae and also was identified in the spectrum of the gas blob ejected by LBV star $\eta$ Car \citep{Zethson_2012}. Further clarification of this issue is desirable, particularly to confirm the correctness of the proposed identification since a stronger line of this multiplet [\ion{Fe}{II}] 6188.55\AA\, with a higher transition probability was not observed.  

\subsection{Proper motions}
\label{sec:PM}

The 13.5 years epoch difference between \textsc{GMOS} and \textsc{SCORPIO} imaging of BP Psc outflow allowed us to confidently detect the proper motions of its knots. Fig.\ref{fig:1} illustrates the positional changes within NE lobe. Displacement of the N1-4 knots is clearly visible in the figure. It is important to note that, within this time span, the general structure of the jet was preserved. We did not detect the appearance of new features or the disappearance of already existed ones. We also found no substantial changes in the internal structure of the knots except those caused by difference in seeing conditions and some contrast effects.

\begin{figure}
	\includegraphics[width=1.0\linewidth]{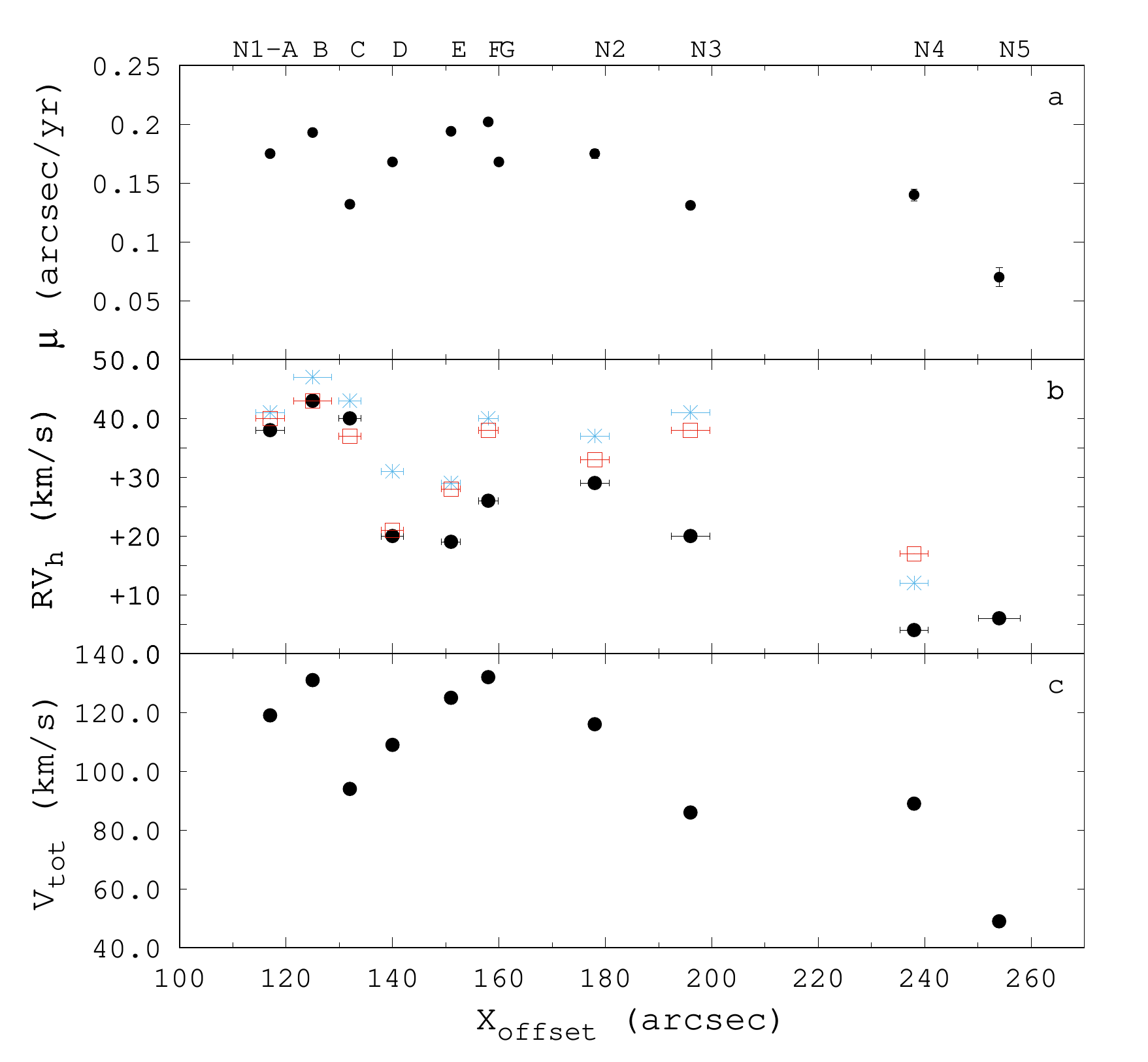}
    \caption{Proper motions (a), radial velocities (b) and spatial velocities (c) in the NE lobe of BP Psc jet as a function of distance from the source. Panel b shows radial velocity measurements with $H\alpha$ (dots), $[\ion{N}{II}]_{6548, 6583}$ (open squares) and $[\ion{S}{II}]_{6716, 6731}$ (asterixes) lines. The horizontal dashes show the spatial limits used for spectrum binning.}
    \label{fig:3}
\end{figure}

The difference in resolution and contrast between images complicate the selection of reference points in the two images for accurate proper motions measurement. Therefore we firstly used the simple model, assuming linear deformation within jet depending on the pixel coordinate. The coordinates transformation between two epochs can be expressed for each point within the jet as $x(t_2) = x(t_1)+\Delta x+k_x x(t_1),\,y(t_2) = y(t_1)+\Delta y + k_y y(t_1)$. Here the transformation constants $\Delta x,\,\Delta y, k_x, k_y$ describe the total shift of a detail (in pixels) and its stretch. We selected a set of the most sharply defined knots that might be easily identified on both images and used their centers of gravity for this linear solution to estimate $\Delta x,\Delta y$. This first approximation solution was used to identify matching details (sets of pixels) in all knots. The images of the knots were approximated using shapelet decomposition \citep{Refregier_2003} in the 36-pixel box. Finally, the photocenters of the knots were determined for each epoch and proper motions $\mu_{\alpha},\mu_{\delta}$ were calculated. The accuracy of measurements was typically better than 0.01$\arcsec \cdot$yr$^{-1}$ for the total proper motion $\mu=\sqrt{\mu_{\alpha}^2+\mu_{\delta}^2}$. Results of our proper motion measurements are summarized in Tab.\ref{tab1} and also plotted in the Fig.\ref{fig:3} for NE outflow.
One can see the general trend of decreasing proper motions with the distance from the star. Indeed, the largest average value $<\mu>=0.176$$\arcsec \cdot$yr$^{-1}$ was detected for the N1 filament at the distances 117-160$\arcsec$ from the source, while a gradual decrease in motion was observed for the more distant knots. The minimum value in the NE lobe $\mu=0.076$$\arcsec \cdot$yr$^{-1}$ was found for the most distant N5 working surface, which also hardly shifted in Fig.\ref{fig:1}. An even smaller proper motion $\mu=0.056$$\arcsec \cdot$yr$^{-1}$ was obtained for S4.

\begin{figure}
	\includegraphics[width=1.0\linewidth]{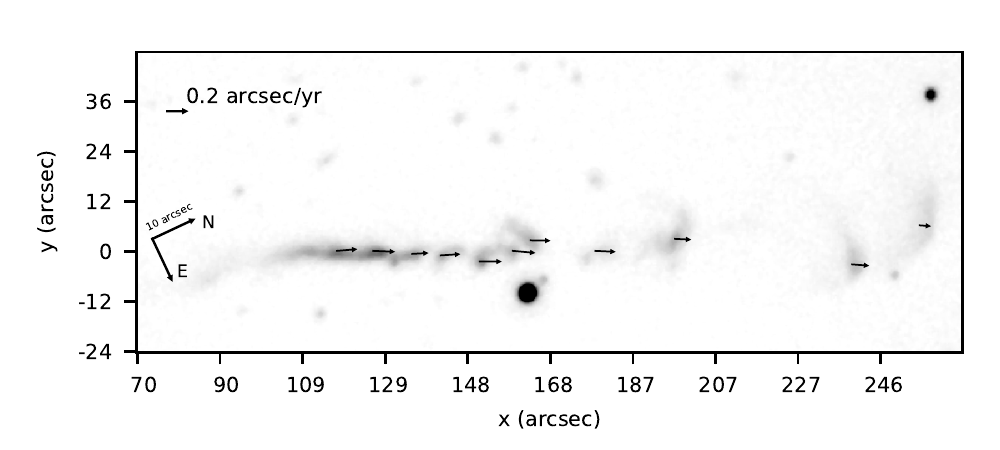}
    \caption{Proper motion vectors overplotted on the SCORPIO image of NE outflow. A scale vector is shown at the top left.}
    \label{fig:4}
\end{figure}

However, a closer examination of the Fig.\ref{fig:3} reveals the existence of a low-amplitude wave-like oscillations superimposed on this general trend. This wave is most prominent within the N1 filament. To clarify the origin of these oscillations: whether they are evidence of the different ejection velocity of the knots, or a consequence of the time-dependence of the ejection direction \citep[e.g.][]{Masciadri_2002,Terquem_1999}, we overplot the proper motion vectors on the image of NE lobe (Fig.\ref{fig:4}). One can see from the figure (and also from Tab.\ref{tab2}) that the position angle of proper motion successively deviates to one side or to another from the outflow axis with average PA=24$^\circ$ that is in agreement with the variability of the outflow direction.

\subsection{Radial velocities}
\label{sec:RV}

Heliocentric radial velocities were measured by Gaussian fitting of emission lines in extracted 1D spectra with \textsc{IRAF} \textit{splot} tool. Results of these measurements for $H\alpha$, [\ion{N}{II}] and [\ion{S}{II}] lines are summarized in Tab.\ref{tab1} and shown in Fig.\ref{fig:3}. In the case of nitrogen and sulphur, the velocities are given as the average value for both component of doublets.

Our radial velocity measurements confirm the orientation of the system suspected on the base of disk geometry in previous studies \citep[Z08,][]{de_Boer_2017}. Indeed, the NE lobe is reciding (redshifted), while the SE lobe is approaching (blueshifted).

As can be inferred from the PV diagram (Fig.\ref{fig:2}) and plot in Fig.\ref{fig:3} the radial velocities within the NE jet decrease with distance from the source. Averaging for N1 filament provides the velocity $<RV(H\alpha)>=+31$ km$\cdot$s$^{-1}$. Within the error the same value was measured also for the N2 knot at the distance 178 $\arcsec$ from the source, while the N3 knot located $\approx20\arcsec$ further possesses lower velocity $RV(H\alpha)=+20$ km$\cdot$s$^{-1}$. The lowest velocities were observed for the most distant N4 and N5 working surfaces: $RV(H\alpha)=+4$ km$\cdot$s$^{-1}$ at 238$\arcsec$ and $RV(H\alpha)=+6$ km$\cdot$s$^{-1}$ at 254$\arcsec$ respectively. The only measured in SW lobe S4 working surface possesses $RV(H\alpha)=-23$ km$\cdot$s$^{-1}$ at 165$\arcsec$. Radial velocities measured for [\ion{N}{II}] and [\ion{S}{II}] lines turn out to be systematically larger than by $H\alpha$, reflecting differences of kinematics of the gas with different excitation conditions.

Relatively low radial velocities together with noticeable tangential motion are consistent with the jet propagation almost in the plane of the sky, in agreement with the BP Psc system's inclination $i\approx70-80^\circ$ \citep[Z08,][]{de_Boer_2017}. A detailed comparison of the tangential and radial velocity components yields an interesting result. As one can see from Fig.\ref{fig:3} the distance dependence of radial velocity within N1-3 knots also possesses the wave-like pattern, but roughly in antiphase with proper motion variations. This effect is most prominent in the N1 filament, where knots D-F show near-maximum proper motions and minimum radial velocities and vice versa for knot C. Such a behavior of proper motions and radial velocities is indeed expected in case of spatial wiggling of the outflow which resulted in variable projection effect for the velocity vector. However, the full spatial velocities $V_{tot}$ also show sinusoidal pattern within N1 filament (Fig.\ref{fig:3}, panel c). Assuming that the knots moving ballistically, this indicates that in addition to the spatial wiggling, there may be also variations in the initial ejection velocity.

In contrast N4-5 working surfaces show decrease in both radial velocities and proper motions which indicates their lower spatial velocity. The average value of the full spatial velocity determined within the N1 filament is $<V_{tot}>\approx118$ km$\cdot$s$^{-1}$, while for N5 this value is only $V_{tot}\approx49$ km$\cdot$s$^{-1}$. Though spatial velocity of knot S4 $V_{tot}=44$ km$\cdot$s$^{-1}$ is close to those of N5, the former has a larger component of velocity along the line of sight, which probably indicate the bending of the SW lobe in the meridional plane.

\subsection{Distance to BP Psc}
\label{sec:Dist}

Prior to the \textsc{Gaia} era, the parallax of BP Psc remained unknown. Indirect estimates of the spectroscopic parallax have been complicated by the uncertainties in the spectral classification of the star. Thus, with an independently determined distance, it would be possible to solve the inverse problem of determining the luminosity class and, hence, the evolutionary status of the star. 

Direct measurements of the BP Psc parallax appeared recently in the second and third \textsc{Gaia} releases \citep{GAIA_DR2,GAIA_EDR3}: $\pi=2.79\pm0.39$ mas and $\pi=5.79\pm0.50$ mas respectively. These parallaxes can be converted into distances: $D=358^{+59}_{-45}$ pc and $D=209^{+24}_{-20}$. Despite a small formal error, the distances from the two data releases differ by a factor $\approx1.7$. Such a large discrepancy is explained by the large values of the $RUWE$ parameter: 10.5 and 12 provided in DR2 and DR3 respectively. Recall that $RUWE$ characterizes the quality of fitting observational data by the astrometric model, and the astrometric solution with $RUWE\lesssim1.4$ should be considered as reliable one\footnote{\url{https://dms.cosmos.esa.int/COSMOS/doc_fetch.php?id=3757412}.}. Thus, the \textsc{Gaia} astrometric solutions for BP Psc probably are untrustworthy. The obvious reason is that BP Psc is observed via the light scattered on its circumstellar environment. Our polarimetric monitoring revealed significant changes in scattered light geometry in BP Psc system on time scales ranging from weeks to several years (in preparation). This may cause a shift of the image photocenter during the positional observations and introduces systematic errors into the astrometric solution. Thus, we still have to rely only on the indirect methods to determine the distance to the star. 

One such possibility for distance estimation is provided by the 3D kinematics of the jet. Indeed, the observed tangential and radial velocity components are related to the distance by a simple expression:
\begin{equation}
\label{eqn:1}
 D=\frac{RV\cdot\tan{i}}{4.74\cdot \mu}
\end{equation}
Here $RV$ is the radial velocity in km$\cdot$s$^{-1}$, $i$ - inclination angle of the jet axis to the line of sight, $\mu$ - total proper motion in $\arcsec \cdot$yr$^{-1}$. The major source of uncertainty here is the $\tan{i}$ factor, so this formula is often used to solve the inverse problem of determining the inclination of a jet with a known distance to the source. However, in the case of BP Psc, we have an advantage since the inclination of its disk is known with high accuracy from direct measurements with high angular resolution (\citetalias{Zuckerman2008}, \citet{de_Boer_2017}). The latter authors gave the $i=78.9^{\circ}$ for the inclination angle of the disk's rotational axis. However, we prefer the more conservative value $i=75\pm5^{\circ}$ \citepalias{Zuckerman2008} because the $\pm 5^\circ$ error budget can also account for the effects of deviation of the outflow axis from the geometric axis of the system. Our estimation of the amplitude of the possible precession wave based on the morphology of the NE jet is of the same order of magnitude, although averaging over a set of knots distributed through the several wave cycles can reasonably compensate this effect. Thus, using the values $<RV>=+29.4$ km$\cdot$s$^{-1}$ and $<\mu>=0.171$ arcsec$\cdot$yr$^{-1}$ averaged for the N1-3 knots and excluding bow shocks N4-5 as well as S4 from the consideration we \textit{estimate} from the Eq.(\ref{eqn:1}) the distance to BP Psc as $D\approx135\pm40$ pc. This distance is less than the values provided by \textsc{Gaia} and indicates that BP Psc is located relatively close to the Sun.

\subsection{Luminosity of BP Psc system and properties of its central source}

With the distance estimate in hand, we can discriminate between the aforementioned scenarios of BP Psc as "isolated YSO"\, or  "post-MS giant"\,. The 135 pc distance unambiguously points to the former. Nevertheless, exact estimation of the central star's luminosity is complicated by number of factors. The major of them is the error in distance. Secondly, according to the high resolution observations \citep[Z08,][]{de_Boer_2017}, the central object isn't observed directly at all, being obscured by nearly edge-on optically thick disk. Instead, we observe scattered light from less obscured parts of the circumstellar environment. The observed flux is a function of stellar luminosity and unknown factors reflecting scattering material's geometry and optical properties. In this section we explain our efforts to estimate stellar luminosity taking in account all relevant effects and uncertainties.       

\begin{figure}
	\includegraphics[width=1.0\linewidth]{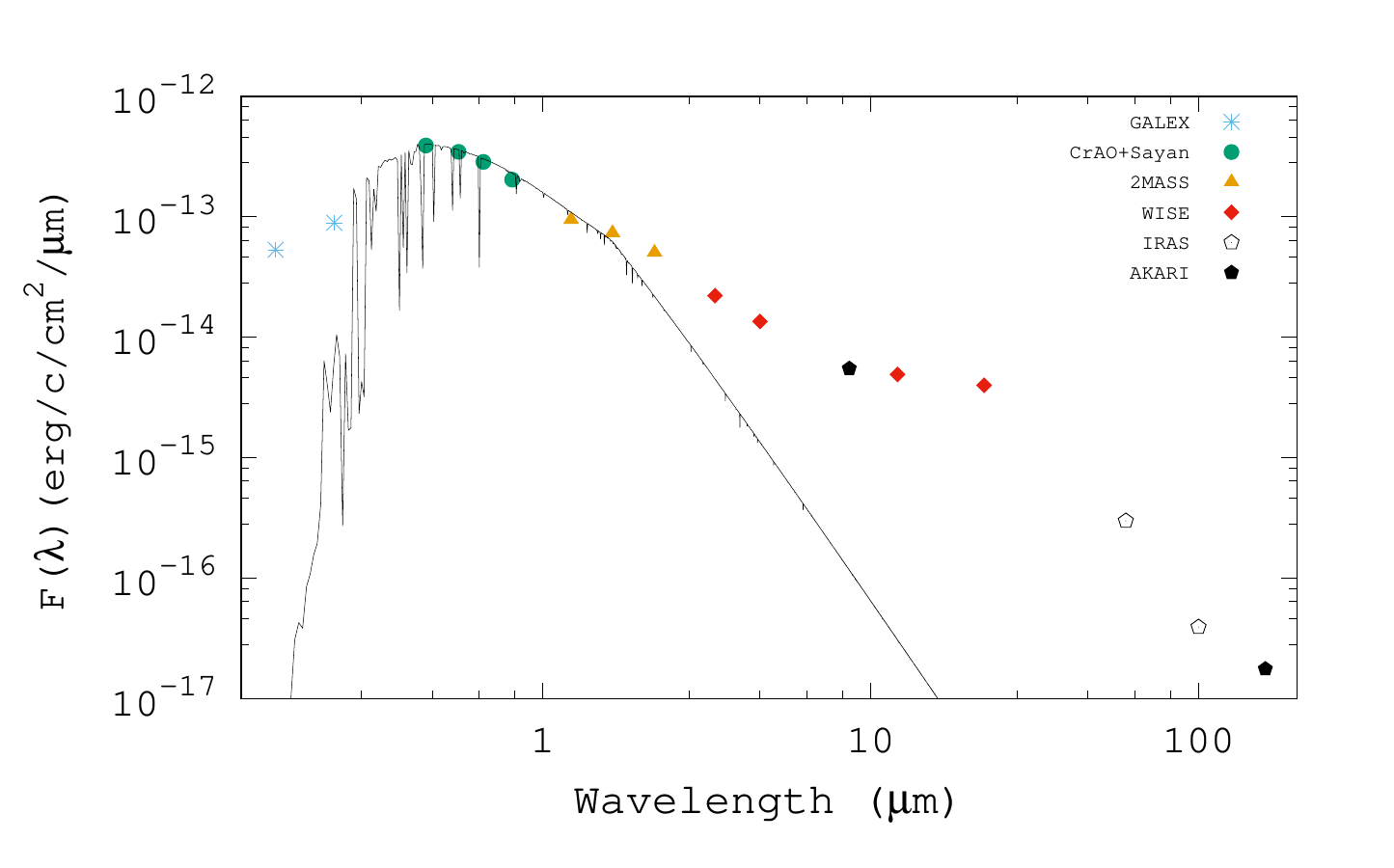}
    \caption{SED of BP Psc. The dots of different types correspond to photometry dereddened with $A_V=1.5^m$, thin line represents synthetic flux for $T_{eff}=5100K$ photosphere.}
    \label{fig:5}
\end{figure}

Previously SED of BP Psc was studied by \citetalias{Zuckerman2008} and \citet{Melis_2010} who showed that at wavelengths longer than $\sim1\mu$m, it is dominated by thermal radiation of circumstellar dust. It was also shown by the authors cited above that IR part of the SED has a double-peaked form which can be fitted with two Planck functions having temperatures 1500K and 210K respectively. We have constructed SED of BP Psc in the $0.15-160 \mu m$ range, using archival UV and IR photometric data from the \textsc{GALEX} \citep{Bianchi_2011},  \textsc{2MASS} \citep{Skrutskie_2006},  \textsc{WISE} \citep{Cutri_2014},  \textsc{AKARI} \citep{Yamamura_2010} surveys and  \textsc{IRAS}-Point Source Catalogue. Optical $BVRI$ magnitudes were obtained as a part of our photopolarimetric monitoring of BP Psc in 2017-2022 at Crimean and Sayan Solar observatories. The magnitudes were averaged and converted to absolute fluxes in Vega system. The resulting SED is plotted in the Fig.\ref{fig:5} and shows a significant excess of emission in the IR region with traceable two peaks near 4 and 25$\mu$m separated by the depression at 7-8$\mu$m. 

The interstellar extinction in the line of sight toward BP Psc is negligibly small as follows from both the 3D dust extinction map \citep{Green_2015} and our observations of interstellar polarization of neighboring field stars which does not exceed $P_{IS}\approx0.5\%$. The low extinction is also indirectly evidenced by the abundance of faint galaxies visible on direct images of this region. Therefore, the total incident flux $F_{obs}$ can be obtained by the direct integration under the SED and luminosity of the "star+disk"\, system can be calculated as $L_{bol}=4\pi D^2 F_{obs}+\Delta L_{\infty}$. Where $\Delta L_{\infty}$ is the bolometric correction to account for luminosity at wavelengths longer than 160$\mu m$, where the observed SED truncates. Using the analytical expression by \citet{Chavarria_1981} we estimated correction to be almost negligible $\Delta L_{\infty}\approx0.02L_{\odot}$, and hence the bolometric luminosity is $L_{bol}\approx2.2\pm1L_{\odot}$ for the adopted distance $D\approx135\pm40$ pc.

In order to isolate the contribution of the central star, we fit the optical part of the SED by the grid of synthetic fluxes calculated for the stellar photosphere with effective temperature within $T_{eff}=4500-5500K$ range. The computations of the emergent flux were performed with SynthV code \citep{Tsymbal_1996} using the atmospheric models interpolated from the MARCS grid \citep{Gustafsson_2008}. Theoretical fluxes were diluted for the 135 pc distance to the source. Together with the temperature the value of the circumstellar extinction, which also accounts effects of scattered light (see explanation below) was found using the \citet{Fitzpatrick_1999} extinction curve and standard reddening law $R_V=3.1$. The reliability of using the interstellar extinction curve is justified by the similarity of the optical characteristics of the circumstellar dust particles to the interstellar ones (in preparation). One can see from the Fig.\ref{fig:5} that theoretical flux with $T_{eff}=5100K$ fits the optical region of observed SED resonably well requiring extinction correction with $A_V=1.5^m$. Here it is worth to give a brief comment on the derived extinction value. At first glance, moderate extinction $A_V=1.5^m$ contradicts to appearance of BP Psc in high-angular resolution images which revealed that the star is deeply embedded in the dusty disk and does not directly visible at optical or NIR wavelengths \citepalias{Zuckerman2008}. We point that obtained $A_V$ is rather formal value which is related to both the direct radiation of the star and scattered light, which dominates in such a case. Thus the "true"\, extinction for the central star can be very large, but the observed flux is specified by the optics of dust scattering.  

From the comparison of the observed flux with photosperic ones it is clearly seen that BP Psc possesses the typical SED of accreting T Tauri star \citep[e.g.][]{Furlan_2006} with both prominent IR and UV excesses. The dereddened optical $BVRI$ points and NIR $J$ and $H$ bands trace the stellar photosphere, while begining from $K_S$ band the thermal dust emission becomes large enough to overwhelm the stellar flux. The origin of excess emission observed in the GALEX NUV and FUW bands most probably related to radiation of the hot region of accretion shock, although a some contribution could be due to the scattered light. 
Integration of the flux under the photospheric model resulted in luminosity of central star $L_*\approx 1.2L_{\odot}$. The obtained values indicate a large contribution of the infrared luminosity of the disk to the bolometric ones $L_{IR}/L_{bol}\approx$80$\%$, in agreement with previous estimation by \citetalias{Zuckerman2008}. In general, the bolometric luminosity of BP Psc system is of order of the luminosities of classical T Tauri stars with accretion disks \citep[e.g.][]{Kenyon_1995}.

Adopting the effective temperature $T_{eff}=5100K$ and using the just found luminosity $L_*$ we place BP Psc on HR diagram and find the mass $M=1.3M_{\odot}$ and age $t\approx7$ Myrs comparing its position with the theoretical evolutionary tracks and isochrones from the PARSEC models \citep{Bressan_2012}. However given the appearance of the jet which is typical for much younger source and complicated procedure of the luminosity estimation we prefer to consider the above age estimation as the upper limit. Even within this uncertainty the present position on HR diagram indicate that BP Psc is still PMS object and its mass lies near the upper mass limit of classical T Tauri stars \citep{Grankin_2016,Villebrun_2019}. On the evolutionary track it locates near the first luminosity minimum caused by development of the radiative zone and transition to the Henyey phase. When it reaches the ZAMS, BP Psc will be star of the $\sim$F5 spectral type. 


\section{Discussion}

\subsection{Evolutionary status of BP Psc system and origin of the jet}

In the absence of distance estimation, the set of unusual observational characteristics of BP Psc was interpreted from the confliting positions, considering it as a young T Tauri star \citep{Downes&Keyes1988,Whitelok1995} with accretion disk or as evolved giant engulfing its planetary system \citep{Zuckerman2008,Melis_2010}. While the recent measurements of parallax of BP Psc by the \textsc{Gaia} mission seem to be unreliable as indicated by the $RUWE$ parameter exceeding significantly the threshold for confident astrometric solution, our estimation based on the radial velocity and proper motion study of the jet leads to a distance $D=135\pm40$ pc. Integration under the SED assuming this distance results in bolometric luminosity $L_{bol}\approx2.2\pm1L_{\odot}$ for the "star+disk"\, system. The best-fit model of the photospheric part of the SED with $T_{eff}=5100K$ photosphere indicates luminosity $L_*\approx1.2L_{\odot}$ for the central source which points unambigously to its dwarf luminosity class. Indeed the position of BP Psc on the HR diagram corresponds to the PMS star with mass $M=1.3M_\odot$ and age $t\approx7$ Myrs - the parameters which are typical for a young T Tauri star.  

The results of our study of the BP Psc outflow show that its physical characteristics strongly resemble those of HH-jets driven by young stars. Indeed, the knots within the jet exhibited a typical low-excitation emission spectrum including $H\alpha$ line and [\ion{N}{II}], [\ion{S}{II}] and [\ion{O}{I}] forbidden doublets. The only tentative identification of somewhat higher excitation emission [\ion{Fe}{II}](44F) was made at distances of several arcminutes from the source. The proper motions measured over a 13.5-years epoch difference combined with the radial velocity measurements resulted in a full spatial velocities up to 140 km$\cdot$s$^{-1}$ for the $H\alpha$-emitting gas. Meanwhile, the outflow is also highly collimated with opening angle $\phi \approx2.2^{\circ}$ measured in NE lobe at the distances up to 2.5$\arcmin$ from the source. According to the modern concept, such a supersonic velocities that are not accompanied by the significant lateral expansion of the jet can only be achieved in the presence of a large-scale magnetic field \citep{Ferreira_2013,Ray_2021} which is naturally expected for the young T Tauri star.

Even without relying on a distance estimation, we can rule out the possibility of the jet launching from a single red giant surrounded by secondary generated disk on the groung of the jet's physical characteristics. Such a disk should not host the magnetic field that supports the launch and collimation of the disk wind. The jet can also be launched from the magnetosphere-disk interface along the open field lines, but the known magnetic fields of single red giants \citep{Auriere_2015} are not sufficient for such a mechanism. Actually, evolved objects like post-AGB stars and symbiotic systems sometimes also possess much focused jets \citep[e.g.][]{Vlemmings_2006,Melnikov_2018} in addition to wide-angle flows driven by the powerful stellar winds. The formation of such outflows is explained by presence of magnetized secondary companion (e.g. white dwarf) or in exotic scenarios of secondary dynamo-generated magnetic field \citep[e.g.][]{Blackman_2004}. We can also rule out these possibilities from the evidences of little resemblance of BP Psc to a symbiotic system. For BP Psc system our estimation of mass-loss rate $\dot{M}_{out}\approx1.2\cdot10^{-8}M_{\odot}\cdot yr^{-1}$ is at least an order of magnitude lower then those in symbiotic systems. Also there are no any traces of wind-driven shells or another signatures of non-collimated mass-loss visible in the direct images in vicinity of the object. The observed SED also does not match the combination of a cool red giant with a hot $T_{eff}\sim10^4$K white dwarf, which should contributes to a more powerful UV-excess.

On this basis, we conclude that young evolutionary status of BP Psc allows to consistently link observational manifestations of its activity with accretion/outflow processes in the primordial disk.

\subsection{Origin of BP Psc at high galactic latitude}

If BP Psc is a PMS star with an age less than 7 Myr, how its isolated position at galactic latitude $b=-57^{\circ}$ can be explained? Without a special study of its kinematics, we can only briefly speculate on this issue. Our distance estimation $D=135\pm40$ pc indicates that BP Psc is a nearby object located almost at the same distance as young stars in the Taurus cloud. Its spatial position in heliocentric Cartesian coordinates is $(X,Y,Z)=(14.5,71.6,-91.7)$pc. Here the 16 pc displacement of the Sun toward the North Galactic pole \citep{Bobylev_2016} was taken into account. Its $Z$ coordinate indicates that BP Psc lies at $\approx92$ pc south of the Galactic plane. According to \citet{Bobylev_2020} at given Galactic longitude $l=78.6^{\circ}$ this value is still within the vertical scaleheight of the young low mass population of the Gould Belt which is tilted with respect to the Galactic equator. The spatial motion of BP Psc $(UVW)_{LSR} = (-10.1;-20.2;5.1)$ km$\cdot$s$^{-1}$ was calculated for adopted distance with \textsc{Gaia} EDR3 astrometric data and the systemic radial velocity $R_V = -17.6$ km$\cdot$s$^{-1}$ \citepalias{Zuckerman2008}. The $UVW$ velocities also corresponds to the kinematics of the thin Galactic disk. 

Overall, spatial position and kinematics do not contradict to BP Psc youth. However, a direct comparison of the $XYZ$ coordinates and $UVW$ velocities does not allow to attribute BP Psc to one of the nearby young moving groups, like $\beta$ Pic, TW Hya, $\eta$ Cha etc. \citep{Zuckerman_2004,Torres_2008}. BP Psc is more distant, lying outside of the Local Bubble, much younger and active than members of this groups with typical age larger than $\sim10$ Myr. Recently the nearby spherical system of molecular clouds named "Per-Tau shell"\, was identified \citep{Bialy_2021} in second Galactic quadrant. In addition to the Taurus and Perseus star-forming regions it also contains a system of diffuse clouds extending to high Galactic latitudes. The origin of "Per-Tau shell"\, was assigned to $\approx6-22$ Myr old Supernova explosions which compressed the diffuse interstellar medium into denser clouds. In turn, these clouds may be sites of subsequent triggered star formation. BP Psc cannot be attributed to the main system of clouds tracing the "Per-Tau shell"\,, mainly due to the difference in the $X$ coordinate. However, it could have formed in one of the peripheral cloudlets. To clarify this issue, a trace-back calculations of the galactic orbit of BP Psc with respect to "Per-Tau shell"\ and search for coeval stars are desirable. Our preliminary searches have shown that there are a few stars within 5$^\circ$ radius around BP Psc sharing similar spatial motion and possessing infrared excesses. Further search including confirmation spectroscopy seems to be promising.

\subsection{Ejection history in BP Psc system}

The characteristic feature of many YSO jets and the BP Psc outflow in particular is their patchy morphology. The most likely explanation for the origin of this structure is based on the model with time-dependent outflow velocity which produces the internal shocks within the flow \citep[e.g.][]{Raga_1990}. If we assume that the knots in the BP Psc outflow are indeed internal shocks produced by the variations in ejection velocity, than their arrangement allows us to trace the history of outflow activity of the star. In our opinion, this assumption is quite reasonable, since the measurements show pronounced variation in spatial velocity both on the scale of the individual N1 filament and within the entire NE lobe of BP Psc outflow (Fig.\ref{fig:3}, panel c). We consider this to be a sufficient argument for the observed pattern of HH objects in BP Psc outflow being formed by the collision of faster clumps with slower traces of previous eruptions, rather than by any kind of plasma instability.

The straightforward application of this method is based on the estimation of the dynamical age $t_{dyn}$ of the knots assuming their ballistic motion. For simplicity, we calculate $t_{dyn}$ based on proper motions: $t_{dyn}= \alpha/\mu$, where $\alpha$ is the angular distance of the knot from the source. The dynamical ages obtained in this way indicate the unsteady character of the star's outflow activity in consistency with the visual impression of irregular distribution of knots along the NE lobe. The last episode of enhanced activity that produced the N1-N2 knots (and probably symmetrical S3) occurred between $\approx$700-1000 yr ago. The N3 and N4 knots were ejected $\approx$1500 and 1700 yr ago respectively. It took  $\approx$3350 yr for the most ancient N5 working surface to travel to its present position, and it is not clear whether the ejection of S4 knot was synchronized with N4 or N5 since it has intermediate age $\approx$2850 yr. The recent MHD simulations \citep[e.g.][]{Romanova_2018} of the magnetospheric accretion/outflow processes has shown their pulsating character. However, such a variability modulates in the region of star-disk interaction on the timescale of few rotational periods of the star. Therefore, it is difficult to accept this mechanism as the only explanation for observed distribution of the knots along the jet, which indicates for example periods of quiescence with duration up to $\sim$1500 yr.

However, as was shown by \citet{Raga_2002} the interpretation of ejection history in YSO jet-driving systems may not be so straightforward. The visible jet structure is traced by shocks, which can be induced by variations of ejection velocity near the source in two (or even three) overlapping modes with different amplitudes. For the HH34 and HH111 outflows the slow mode with a period of about $\sim10^3$ yr and a fast mode on the scale of decades and with a smaller amplitude of velocity changes was found. It has also been shown that the maxima of the slow mode can amplify the shocks in the fast mode, forming bright chains of the knots. 

Qualitatively, character of the BP Psc outflow is consistent with the model with multi-mode variability of ejection velocity. It is possible to distinguish a fast low-amplitude variability producing the knotty structure of the N1 filament and suspect large-amplitude variability that modulated formation of the slower bow-shocks N4-5. We estimate the fast mode period as $\approx$54 yr (see the next section) and attribute it to the tidal effects from the secondary companion. However, our exercise with the analytical treatment by \citet[][]{Raga_2015} did not lead to a self-consistent result in determining the period of the slow mode. This is probably due to incorrect attribution of the N4-5 knots as its first working surfaces, or due to existence of another variability mode. We can only estimate the order of magnitude of the slow mode period also as $\sim10^3$ yr. For the further clarification it is desirable to measure velocities of the most distant knots N6-7 and S5-6 as well as to search for others remnants of ancient activity within a larger field of view.

\subsection{Wiggling and asymmetry of the jet as indicators for secondary companion}

Does BP Psc system host a low-mass companion? Its existence was suspected by \citetalias{Zuckerman2008} from the variability of the radial velocities measured in high-resolution spectra. However, this RV variability can also be caused by deformations of the line profiles due to  effects of scattering stellar radiation on the moving circumstellar dust grains \citep{Grinin_2006}. Nevertheless, other observational evidences also point to the possible existence of a secondary companion. First, one possible interpretation of the two-humped SED implies the existence of a gap in the disk, probably opened by the orbital motion of the companion \citep{Melis_2010}. Second, "wiggling"\, jet morphology is also a typical observational manifestation of the variable orientation of the outflow direction which can be caused either by orbital motion in a binary system \citep{Masciadri_2002} or by the precession of the jet-driving disk under the tidal interaction with a companion on a non-coplanar orbit \citep{Terquem_1999}. Observations of BP Psc with SPHERE/ZIMPOL did not reveal presence of the external companion in intensity images within the 1.2$\arcsec$ field of view \citep{de_Boer_2017}, that translates to $\approx$160 au at adopted 135 pc distance. Thus, we assume that the companion, if exists, is located on a closer orbit inside the circumbinary disk.

In this case, the underlying physics is similar to those in \citet{Terquem_1999} scenario, with the difference that only the inner part of the disk, which is broken by the gap, precesses \citep{Zhu_2019}. The latter author gives an expression (F.27) that relates the observables: precession frequency $\omega_p$ and the precession angle $\delta$ with the mass ratio of the component $q=M_p/M_*$. In case of a low-mass ($M_p\ll M_*$) component, this expression can be rewritten as:

\begin{equation}
\label{eqn:2}
 q=\frac{8}{3}\cos \delta \frac{\omega_p}{(GM_*/R_d^3)^{1/2}}
\end{equation}

Here we also assume that the component is orbiting on the inner edge of the cavity and the radius of its orbit coincides with the outer radius of precessing disk, i.e. $R_p\approx R_d$. 
Actually, for a fixed mass of the primary $M_*$, Eq.(\ref{eqn:2}) gives a family of solutions depending on $R_d$ (or, identically, $R_p$) and precession parameters, which can be found from observations. For BP Psc system we adopt mass of the primary $M_*\approx1.3M_\odot$. The deprojected "length"\, of the precession wave within N1 filament was estimated by measuring the positions of few successive maxima on the contour plot as $\lambda=\lambda_{proj}/\sin i = 10.1\arcsec$. This angular value corresponds to a spatial size of $\approx$1360 au at 135 pc distance. Hence, the precession period can be calculated as $T=\lambda/<V>$ and it is of about 54.4 yr, using the average value $<V_{tot}>\approx118$ km$\cdot$s$^{-1}$ of the spatial velocity within the N1 filament. The precession frequency related with the period as $\omega_p=1/T$. Finally, the precession angle measured from the NE jet $\delta = 5^\circ$ leads to $\cos \delta \approx1$.

Is there any way to determine the most plausible range for $R_d$ in our case? BP Psc is photometrically variable with a large-scale brightness dimmings caused by variable circumstellar extinction (in preparation). The Lomb-Scargle periodogram computed on the base of the historical light curve covering $\approx$20 yr revealed the existence of the $\sim1100^d$ period. Assuming that this period is modulated by the motion of a companion on the circular Keplerian orbit, we can estimate the radius of its orbit as $R_d\approx2.3$ au adopting mass of the primary $M_*\approx1.3M_\odot$. Within the uncertainty of stellar parameters, this estimation of the gap radius is consistent with the deficit of the dust with temperature $\sim500-800$K inferred from SED. Substituting these $R_d$ into Eq.(\ref{eqn:2}) results in a mass of the companion $M_p\approx 0.03M_\odot \approx 30M_{Jup}$.

The presence of such a substellar or at least planetary-mass companion on a misaligned orbit can also be responsible for an obseved hemispherical asymmetry of the BP Psc outflow. In fact, the difference between the approaching and receding lobes of YSO outflows is frequently observed \citep[e.g.][]{Melnikov_2009,Gomez_2013} and can be explained either by a real difference in mass-loss rate or by a difference in the excitation conditions. The latter can be caused, in particular, by the pressure gradient of the external ambient medium \citep{Matsakos_2012}. We can practically exclude the difference in environmental conditions in case of BP Psc outflow, because the star does not assotiate with any dense molecular cloud and radio observations revealed the distribution of observed CO emission in form of the molecular disk coincided with its optical counterparts and with no traces of extended envelope \citepalias{Zuckerman2008}. Thus, more likely that different appearance of the NE and SW lobes of the BP Psc outflow is caused by the difference in mass-loss rate or different efficiency in generation of internal shocks. The reason may be both the effects of interaction between the stellar magnetosphere and the magnetic field of the disk \citep{Matsakos_2012,Dyda_2015}, and the geometric effects of shielding by the warped inner disk \citep{Soker_2005}. The disk warping may be caused by the presence of a substellar companion on a non-complanar orbit \citep[see e.g.][and references therein]{Demidova_2013,Zhu_2019}. Further observations of BP Psc, in particular more complete spectroscopy of the SW lobe, are needed to choose between these possibilities.

\section{Conclusions}

We present results of the proper motion and spectroscopic study of the bipolar outflow driven by the high-latitude star BP Psc with unclear evolutionary status. Our measurements on the 13.5 yr baseline reveal the proper motions of the jet knots within $\sim 0.08-0.2$ $\arcsec \cdot$yr$^{-1}$ range. The corresponding radial velocities measured in the redshifted NE lobe are $RV(H\alpha)\approx +4-40$ km$\cdot$s$^{-1}$. Both proper motions and radial velocities show variations within $\sim200^{\arcsec}$ distance from the source indicative for the spatial wiggling and a tendency to decrease for the more distant knots. We estimated the electron density in the knots as $N_e\sim 10^{2}$ cm$^{-3}$ which is typical for low-excitation YSO jets. 

The 3D space motion of the outflow with advantage of the previous accurate measurements of the inclination of its axis ($i\approx75\pm5$) provides the distance $D=135\pm40$. Fitting the observed SED by the synthetic flux for this distance resulted in temperature $T_{eff}\approx5100K$ and luminosity $L_*\approx1.2L_{\odot}$ for the central source of BP Psc system. These parameters on HR diagram correspond to a PMS star with the mass $M_*\approx1.3M_{\odot}$ and age of about 7 Myrs. Given the previously reported accretion activity and lithium overabundance in its atmosphere, this confirms the original classification of BP Psc as young T Tauri star.  

We conclude that structure, kinematics and physical parameters along the outflow are consistent with magnetocentrifugal mechanism of its launching and collimation, that is typical for TTS with accretion disks. The non-stationary character of the ejecta and prominent wiggling of the jet indicate the possible existence of a substellar companion on a non-coplanar orbit. Assuming that the time dependence of the ejection direction is caused by the precession of the inner disk inside the companion's orbit, we estimated its most plausible mass as $\approx$30 $M_{Jup}$. The impact of the secondary companion on the structure of the inner disk and tidal modulation of the ejection velocity can also be the reason for the observed NE-SW brightness asymmetry of the jet.

Given its distance, BP Psc appears to be one of the closest to Earth jet-driving system. Its orientation almost in the plane of the sky and low extinction in this direction makes it an extremely promising object for a detailed study of the jet structure and wind/jet interface region with modern high-angular-resolution and spectropolarimetric facilities.  
The origin of BP Psc is possibly related to the recent star-formation event triggered by the expanding super-shells in the second Galactic quadrant, but this issue requires further investigation as well as the search for coeval objects.

\section*{Acknowledgements}
This research was financially supported by the Ministry of Science and Higher Education of the Russian Federation.\\
Observational data obtained with the telescopes of the SAO RAS were used. Observations with the SAO RAS telescopes are supported by the Ministry of Science and Higher Education of the Russian Federation. The renovation of telescope equipment is currently provided within the national project "Science and Universities".\\
The observations at Sayan Solar Observatory were obtained using the equipment of the Angara Center for Collective Use http://ckp-rf.ru/ckp/3056/.\\
Based in part on observations obtained at the international Gemini Observatory.\\
D. Shakhovskoy's work was supported by the grant \textnumero075-15-2020-780 from the Ministry of Science and Higher Education of the Russian Federation.

\section*{DATA AVAILABILITY}

The data underlying this article will be shared on reasonable request to the corresponding author.




\bibliographystyle{mnras}
\bibliography{reference.bib} 



\bsp	
\label{lastpage}
\end{document}